\documentstyle[twocolumn,prl,aps]{revtex}           

\begin{document}
\draft

\twocolumn[\hsize\textwidth\columnwidth\hsize\csname 
@twocolumnfalse\endcsname                            


\title{Diffusion on a stepped substrate}

\author{Juha Merikoski\cite{PermAddr} and See-Chen Ying}

\address{Department of Physics, Brown University,
Box 1843, Providence R.I. 02912}

\date{\today}

\maketitle


\begin{abstract}

We present results for collective diffusion of adatoms on a stepped 
substrate with a submonolayer coverage.  We study the combined effect 
of the additional binding at step edge, the Schwoebel barrier, the 
enhanced diffusion along step edges, and the finite coverage on 
diffusion as a function of step density.  In particular, we examine 
the crossover from step--dominated diffusion at high step density to 
terrace--dominated behavior at low step density in a lattice--gas 
model using analytical Green's function techniques and Monte Carlo 
simulations.  The influence of steps on diffusion is shown to be
more pronounced than previously anticipated.
\end{abstract}


\vskip2pc]                                       


The importance of surface steps in dynamical processes
such as growth and ordering is widely recognized 
\cite{GrowthOrdering}.
Among other effects, there exist extra activation barriers at the
steps leading to a different mobility for adsorbates there
as compared to the flat terrace region \cite{Sch69}.
In recent years, several new techniques have been employed to measure 
the diffusion rate of adsorbate on well--defined low index
surfaces \cite{Gom90,Xia92,Xia93,Xia95}.
However, since most of these experiments involve studies of density
fluctuations over a length scale much larger than the average distance
between steps on the surface, 
it is important to ascertain whether the measured diffusion constant 
corresponds to the intrinsic value for the wide terrace region, 
or whether it is strongly influenced by the rates near the step edges.
While many theoretical studies on the extra barriers at step edges exist, 
only a few consider the effect of steps on measurable
diffusion constants \cite{Nat93,Ueb94},
and criteria for the crossover from step--dominated to terrace--dominated 
diffusion with increasing terrace width are not well 
established \cite{Xia92}.

In this Letter, we present a study of the effect of various microscopic
modifications of adatom jump rates due to surface steps
on macroscopically observable collective diffusion.
We describe the system of adsorbates on a substrate
with a periodic array of steps by a lattice--gas model.
We use a combination of analytical approach and Monte Carlo simulations 
to analyze the dependence of diffusion on step density in this model.
We discuss the implications of our results on the experimental
determination of diffusion constants, and
the recent measurements by Xiao {\it et al.}\ \cite{Xia92,Xia93,Xia95}
using optical diffraction techniques in particular.

\begin{figure}[h]
\null
\vspace{9.6cm}
\includegraphics{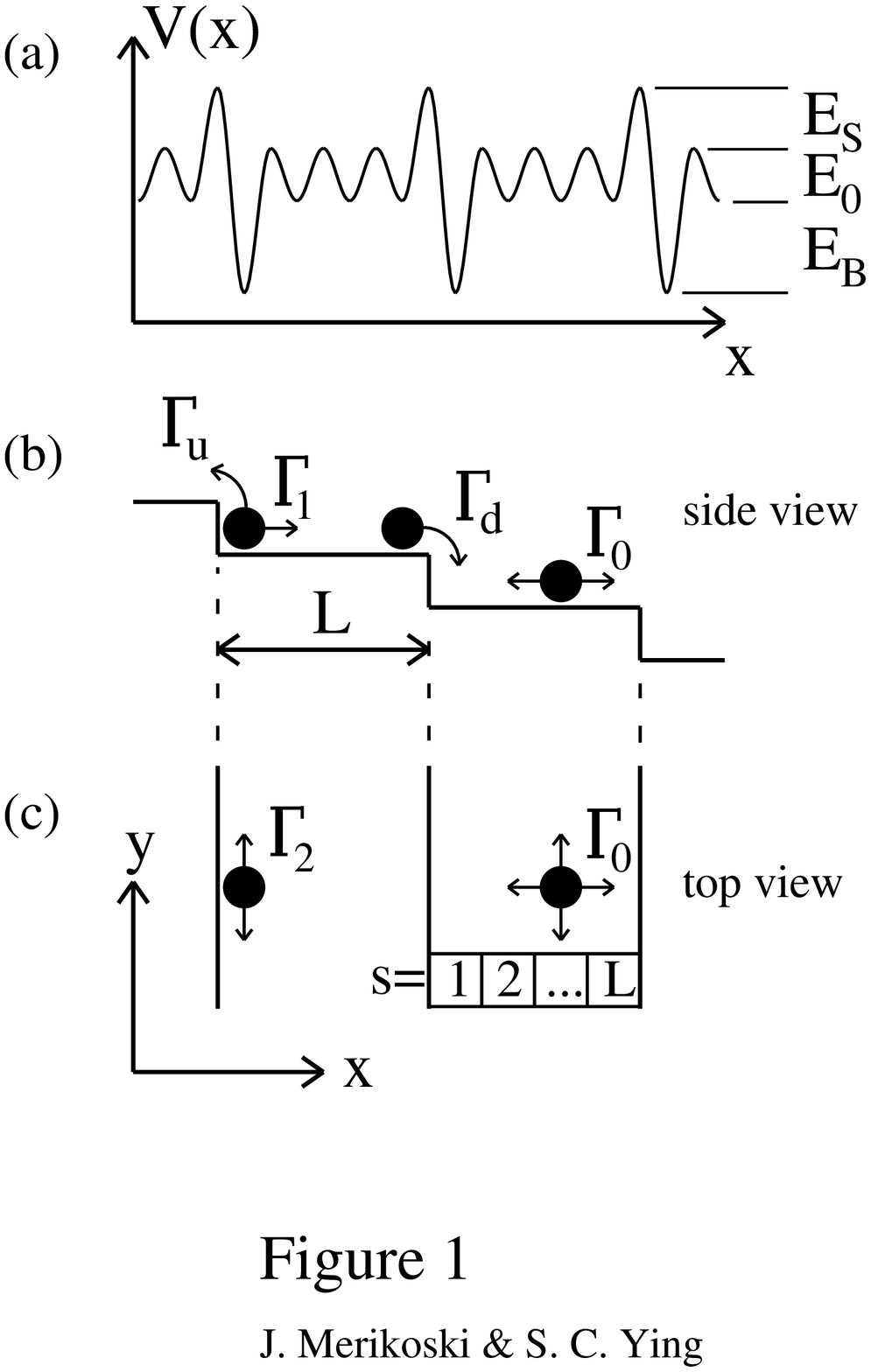}
\caption{Geometry and jump rates of the lattice model for diffusion
on a stepped substrate.
(a) The potential profile in $x$ direction.
(b) Side view of the model showing the various hopping rates
for jumps in $x$ direction near step edges.
(c) Top view of the model showing
the size of one unit cell with the indices $s=1,2,\ldots,L$
of each lattice site within the cell.
}
\label{fig:model}
\end{figure}

Following an earlier work dealing with the zero coverage 
limit~\cite{Nat93},
we introduce several energy barriers
in our model characterizing the adsorption on the  stepped substrate.
These are the additional binding energy $E_B$ at lower step edge,
the Schwoebel barrier $E_S$ for motion across step edges,
the diffusion barrier on terraces $E_0$, and
the barrier for jumps along lower step edges $E_2$ \cite{Sch69}.
The corresponding potential profile in direction perpendicular
to step edges ($x$ direction) is shown in Fig.~\ref{fig:model}(a).
These energy barriers lead to the following jump rates 
as shown in Fig.~\ref{fig:model}(b) and Fig.~\ref{fig:model}(c)
\begin{eqnarray}
\label{eq:rates}
 \Gamma_0 &=& e^{-E_0/kT} = \Gamma_1 e^{E_B/kT} \nonumber\\
 \Gamma_d &=& \Gamma_0 e^{-E_S/kT} = \Gamma_u e^{E_B/kT} \\
 \Gamma_2 &=& e^{-E_2/kT}.\nonumber
\end{eqnarray}
%


The substrate surface has a periodic array of steps separated by
terraces of width $L$ lattice sites [see Fig.~\ref{fig:model}(c)].
Thus each adsorption site in the lattice  gas model can be labelled 
by the
coordinate $(x,y)$ of the unit cell together with a site
index $s=1,2,\ldots,L$ within the unit cell.
We assume hard--core interactions between the adatoms.
For each lattice site
we then define a stochastic occupation variable $n^s_{xy} (t)$,
which can take only the values 0 and 1.
To analyze collective diffusion, we define the Green's functions
$ G^{ss'}_{xyx'y'}(t) = -2\pi i \theta (t)
	\langle u^s_{xy} (t) u^{s'}_{x'y'} (0) \rangle$,
where $s,s' = 1,2,\ldots L$,
the fluctuation variables $u^s_{xy} (t)$ are defined 
by $u^s_{xy} (t) = n^s_{xy} (t) - \langle n_{xy}^s \rangle$, and 
$\theta (t)$ is the Heaviside step function 
\cite{Tah83,Ala91,CollectiveGreen}.

The equations of motion for $G^{ss'}_{xyx'y'}(t)$ follow from the
rate equations for $n^s_{xy}$ \cite{Mer96}.
After Fourier transforming with respect to the time and cell 
coordinates,
we can obtain a closed set of equations for 
the Green's functions $G^{ss'}_{{\bf k}}(\omega)$  in ($ {\bf 
k},\omega$) 
space by decoupling the higher order Green's functions.
In this mean--field approximation, $G^{ss'}_{{\bf k}}(\omega)$
turn out to be inversely
proportional to the determinant of an almost tridiagonal $L \times L$
matrix $A({\bf k},\omega)$.
The interior elements of $A$,
\begin{eqnarray}
\label{eq:Amiddle}
 A_{ss} &=& i\omega +2 \Gamma_0 \cos k_y b -4 \Gamma_0           
\nonumber\\
 A_{s,s\pm 1} &=& \Gamma_0,
\end{eqnarray}
are related to terrace jumps, and the modified elements at the
corners of $A$,
\begin{eqnarray}
\label{eq:Acorner}
 A_{11} &=& i\omega \!+\! 2 \Gamma_2 \cos k_y b
    -\!\! \Gamma_1 \!-\!\! \Gamma_u\! -\!\! 2 \Gamma_2
    -\!(\Gamma_0 \!-\!\! \Gamma_1
    \!+\!\! \Gamma_d \!-\!\!\Gamma_u) c_t\nonumber\\
 A_{22} &=& i\omega + 2 \Gamma_0 \cos k_y b
    - 4 \Gamma_0 - (\Gamma_1 - \Gamma_0) c_e                     
\nonumber\\
 A_{LL} &=& i\omega + 2 \Gamma_0 \cos k_y b
    - 3 \Gamma_0 - \Gamma_d - (\Gamma_u - \Gamma_d)c_e           
\nonumber\\
 A_{12} &=& \Gamma_0 + (\Gamma_1 - \Gamma_0) c_e                 
\nonumber\\
 A_{21} &=& \Gamma_1 + (\Gamma_0 - \Gamma_1) c_t                 
\nonumber\\
 A_{1L} &=& \lbrack \Gamma_d + (\Gamma_u - \Gamma_d) c_e \rbrack
	     e^{-i L k_x a}                                      \nonumber\\
 A_{L1} &=& \lbrack \Gamma_u + (\Gamma_d - \Gamma_u) c_t \rbrack
	     e^{+i L k_x a},
\end{eqnarray}
are related to jumps near step edges.
All the other elements of $A$ are identically zero.
Here $a$ and $b$ are the nearest--neighbour distances in $x$ and $y$
directions, respectively.
By symmetry, the average occupation numbers $\langle n^s_{xy} \rangle$
depend only on the row index $s$.
In the present model, we have only two distinct row 
coverages $c_e$ and $c_t$, 
defined by $c_e = \langle n^1_{xy} \rangle$ for the lower edge row
and $c_t = \langle n^2_{xy} \rangle = \ldots = \langle n^L_{xy} \rangle$
for terraces.
The value of total coverage $c = c_t + (c_e-c_t)/L$ together with
the detailed balance condition 
\begin{equation}
\label{eq:detbal}
 \frac{c_e(1-c_t)}{c_t(1-c_e)} = e^{E_B/kT}
\end{equation}
completely determine the values of $c_e$ and $c_t$.


The diffusion tensor $D$ can be extracted from the diffusive poles of
the Green's functions, i.e.\ from the zeros of $\det A$
in the hydrodynamic limit ${\bf k} \to 0$ and $\omega \to 0$
at $\omega = -i \, {\bf k} \!\cdot\! D \!\cdot\! {\bf k}$.
By expanding the determinant and collecting
the leading terms for small $\omega$ and ${\bf k}$, 
we find the diagonal elements of $D$ to be 
\cite{Mer96} 
\begin{eqnarray}
\label{eq:DxxDyy}
 D_{xx} &=& \frac{(1-c_e)(1-c_t)c_t\Gamma_0 L^2 a^2}
 {\kappa\, \lbrack (L-2)(1-c_e)+(1-c_t)(1+e^{E_S/kT}) \rbrack}
 \nonumber\\
 D_{yy} &=& \frac{1}{\kappa}
 \lbrack (L-1)(1-c_t)c_t\Gamma_0+(1-c_e)c_e\Gamma_2\rbrack b^2 
\nonumber\\
 \kappa &=& (L-1)(1-c_t)c_t + (1-c_e)c_e.
\end{eqnarray}
By symmetry, the nondiagonal elements of $D$ are identically zero.

\begin{figure}[h]
\null
\vspace{9.0cm}
\includegraphics{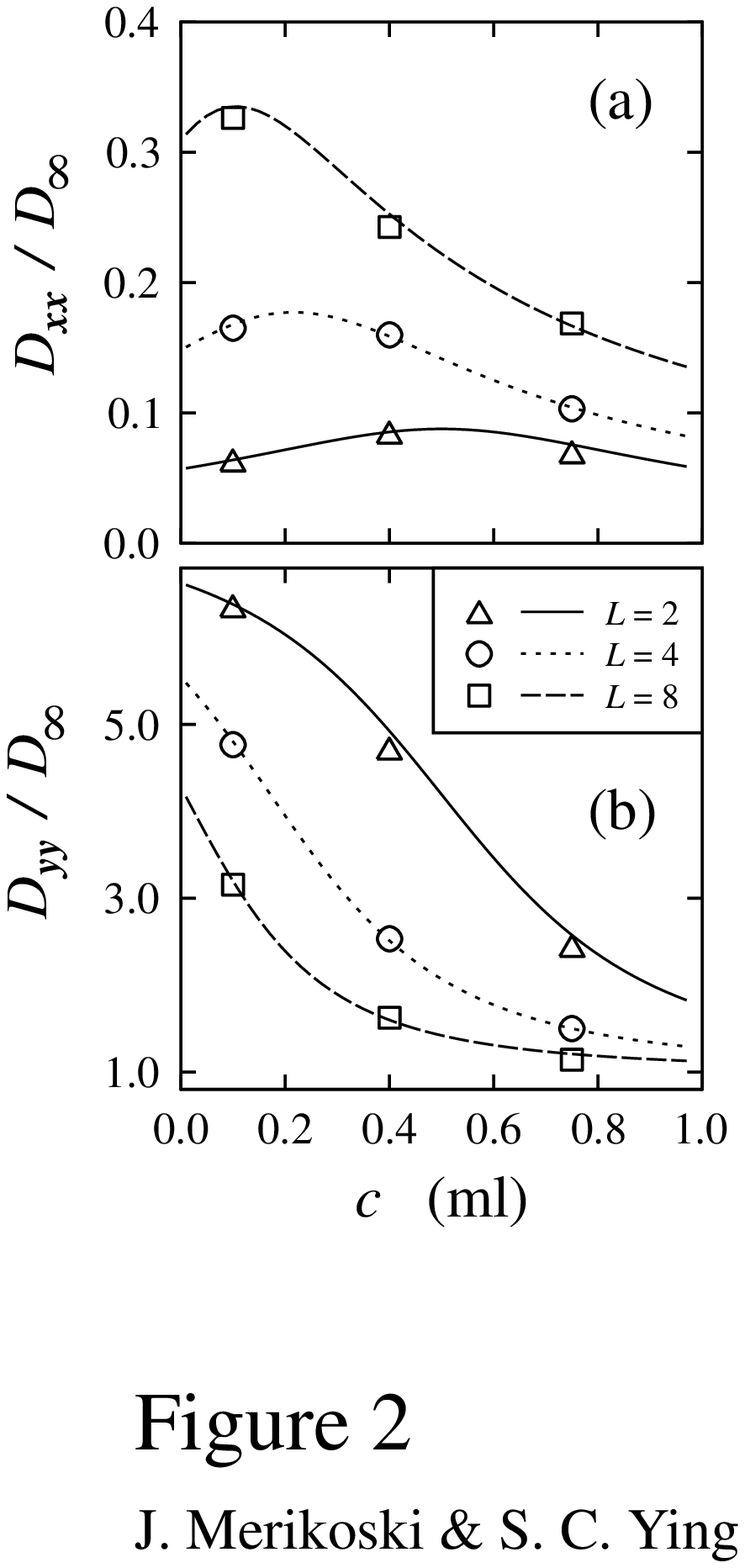}
\caption{Mean field and Monte Carlo results for
$D_{xx}$ and $D_{yy}$ as a function of coverage
at temperature $kT = 1/2$ for the hard--core model
with $E_S = E_B = E_0 - E_2 = 1$.
Lines show the mean field solution of Eq.~(\protect\ref{eq:DxxDyy}), 
and
plotting symbols denote the results of Monte Carlo simulations
with errorbars less than the size of the symbols.
Here diffusion coefficients are shown
in units of $D_{\infty}$ that is the value for infinite
terrace width and coverage in units of one monolayer.
}
\label{fig:mfcov}
\end{figure}

To test the mean--field result of Eq.~\ref{eq:DxxDyy}, we have
performed standard Monte Carlo (MC) simulations \cite{Keh84}
for the same model.
In the MC simulations,
the collective diffusion constant has been calculated
using the Green--Kubo relation \cite{Gom90}
\begin{equation}
\label{eq:GreenKubo}
 D_{\nu\nu} = f
	      \lim_{t\to\infty} \frac{1}{4Nt}
	      \Biggl\langle \Biggl\lbrack
	      \sum_{i=1}^{N} \bigl\lbrack r^{(i)}_{\nu}(t)
				   -r^{(i)}_{\nu}(0) \bigr\rbrack
	      \Biggr\rbrack^2 \Biggr\rangle,
\end{equation}
where the sum is over all particles and $\nu = x,y$.  For our model, 
the `thermodynamical factor' $f$, which is inversely proportional to 
particle number fluctuations \cite{Gom90}, can be calculated exactly 
yielding the result $f = Lc/\kappa$ with $\kappa$ defined in 
Eq.~(\ref{eq:DxxDyy}) \cite{Mer96}.  
The coverage dependence of $D$ as 
given by Eq.~(\ref{eq:DxxDyy}) together with the simulation results 
are shown in Fig.~\ref{fig:mfcov} for a temperature comparable to the 
various activation barriers in the system: 
$kT = 1/2 $ with $E_B = E_S = E_0 - E_2 = 1$.  
Typical simulation parameters were $10^8$ MC steps per atom for 
thousand atoms in the system.
We conclude that the mean--field theory agrees very 
well with the MC result, with no discernible systematic 
deviation \cite{ExactLimits}. 
At low temperatures $f$ has a maximum at $c \approx 1/L$ 
due to the suppression of fluctuations when lower edge 
rows get filled ($c_e\approx 1$)
and terraces are almost empty ($c_t \approx 0$),
resulting in a maximum of $D_{xx}$ at $c \approx 1/L$.
In the case of $D_{yy}$,
this effect of reduced fluctuations is cancelled out by the enhanced
blocking of jumps along the lower step edges.

We now focus on the dependence of the diffusion constant on the width 
of the
terrace as given by the expression in Eq.~(\ref{eq:DxxDyy}).
In this regard,
$D_{yy}$ has a relatively weak dependence on the width,
since diffusion in the direction  parallel to the step edge has to 
proceed through the terrace region as soon as the step edge row is filled. 
Below we shall concentrate on $D_{xx}$,
i.e.\ diffusion perpendicular to the steps.
At very high temperature or for very wide terraces,
the diffusion constant $D_{xx}$ would approach the 
value $D_{\infty}=\Gamma_0 a^2 = e^{-E_0/kT} a^2$ 
appropriate for a single terrace,
and the effect of the steps vanishes.
In the other limit of low temperature and narrow terrace width, 
diffusion is dominated by the steps, and $D_{xx} \propto 
e^{E_{\rm sd}/kT}$, 
where the total barrier $E_{\rm sd}$ is now $E_0 + E_S + E_B$.
The crucial question is how the system crosses over from one limit to 
another.
The crossover behaviour can be understood most easily in the
limit $e^{E_B/kT} \gg 1 $ and $c \gg 1/L $ \cite{OtherLimits},
in which the expression of $D_{xx}$ in Eq.~(\ref{eq:DxxDyy}) 
simplifies to the form
\begin{equation}
\label{eq:Dxxc}
 D_{xx}^{\ast}  = \frac{\Gamma_0 L a^2}
         {L + c (1+e^{E_S/kT}) e^{E_B/kT}},
\end{equation}
from which we see immediately that the crossover occurs at
$L \approx c e^{(E_B+E_S)/kT}$.
For large values of $(E_B+E_S)/kT$,
this implies that the effect of steps could be substantial even for 
terrace widths up to several hundred lattice spacings.

Recently the diffusion of CO on clean Ni(110) was studied
by Xiao {\it et al.} \cite{Xia92,Xia93}. From a measurement on a 
sample with high step densities \cite{Xia95}, they deduce the 
value of the step--dominated barrier $E_{\rm sd}$ for diffusion 
to be 239 meV.
The activation barrier for CO diffusion along (001) on a good
sample with average step separation of $L \approx 170$ was 
determined to be 120 meV.
For diffusion along $(1 \bar{1} 0 )$, a barrier of 95 meV was 
measured.
These authors then use a heuristic argument to arrive at the following
crossover criterion: $L^2 \approx e^{(E_B+E_S)/kT}$. 
Based on this formula they conclude that in the experimental 
temperature 
range of 100 to 200 K,
the influence of the steps should be negligible \cite{Xia92}.
The measured value of 120 meV is then assigned to be the value 
of the barrier $E_0$ appropriate for CO diffusion along (001) on
clean Ni(110).

\begin{figure}[h]
\null
\vspace{4.0cm}
\includegraphics{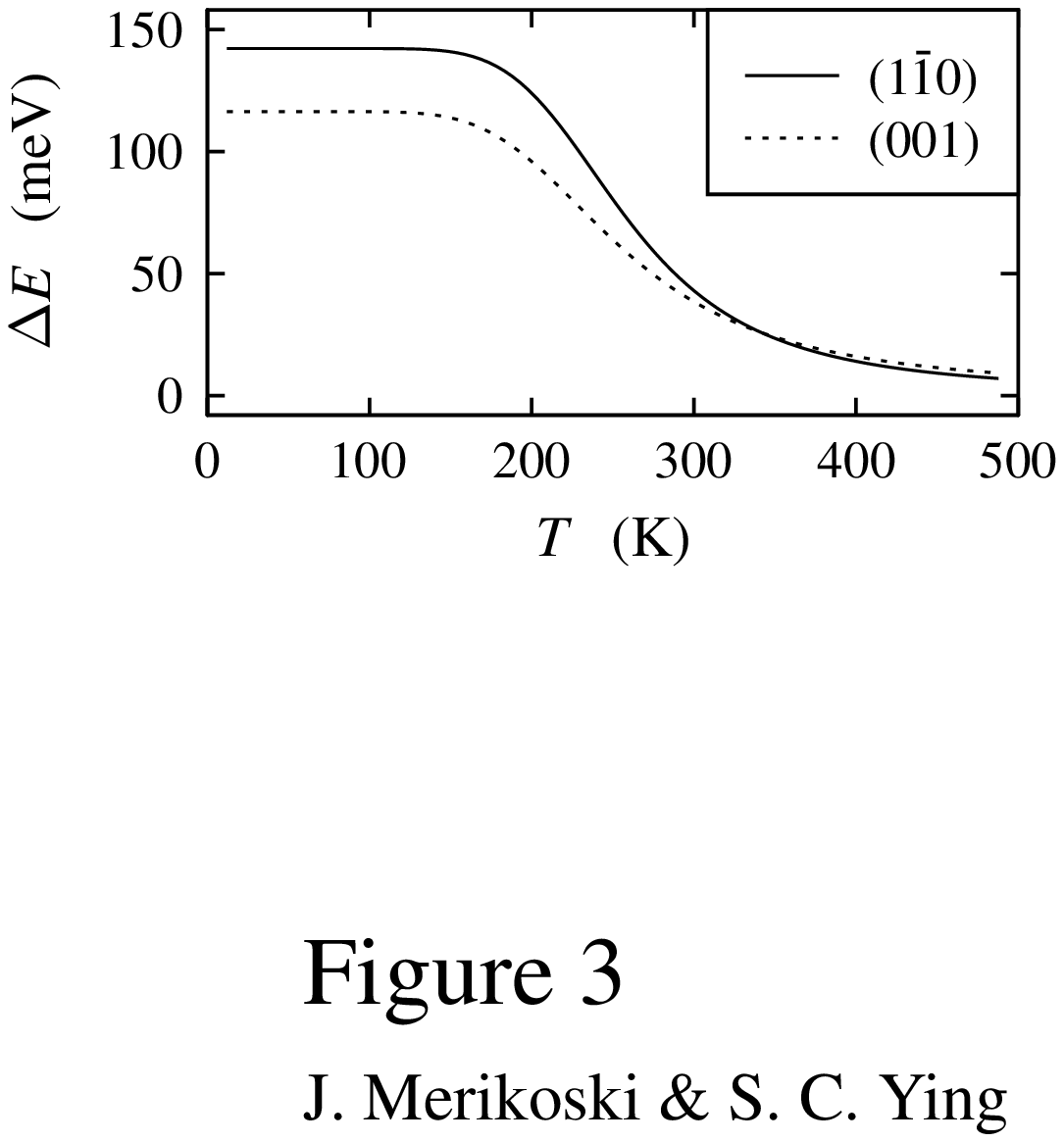}
\caption{The effective extra activation barrier due to the steps
for diffusion of CO along the $( 0 0 1 )$ and $( 1 \bar{1} 0)$
directions on Ni(110). See text for parametrization.
}
\label{fig:CONi}
\end{figure}

We now apply our results to analyze whether the effect of steps is 
indeed negligible under these conditions in this system.  We take as 
input the measured value for $E_{\rm sd}$ of 239 meV and the estimated 
values of $L=170$, $E_0=120$ meV for the (001) direction and $L=520$, 
$E_0=95$ meV for the $(1 \bar{1} 0 )$ direction, and set $E_S = 0$ 
\cite{FeiPC}.  For simplicity, we have also taken all the prefactors 
of the various jump rates to be equal \cite{Prefactors}.  We calculate 
the {\it effective} extra barrier due to the steps, $\Delta E(T,L)$, 
for diffusion perpendicular to the steps, defined as the local slope 
of the Arrhenius plot of $D_{xx}/D_\infty$ using Eq.~(\ref{eq:DxxDyy}) 
for coverage $c=0.5$.  In Fig.~\ref{fig:CONi} we show the result as a 
function of temperature for diffusion along the two directions.  In 
both cases, below 300 K, $\Delta E$ is sizable and the effective 
barrier is not just the terrace value $E_0$ but considerably affected 
by steps.  Clearly the heuristic crossover criterion in 
Ref.~\cite{Xia92} underestimates the influence of the steps.

In a more recent experiment \cite{Xia95}, the effect 
of sulphur as an impurity on diffusion of CO on Ni(110) was studied.
It was proposed that the main effect of sulphur is to increase the  
effective step barrier dramatically. 
A barrier of 323 meV was measured for diffusion along both the main 
orientations.
These authors conclude that in this case, 
the diffusion is dominated by the  sulphur--poisoned step edges, 
and the measured value of 323 meV is then assigned to be $E_{\rm sd}$, 
i.e. the step--dominated barrier.
We have also applied our analysis to this system and find that here 
with $L=520$ assuming $E_S=0$,  the steps indeed 
dominate for $E_0 \le 200$ meV.
So the determination of the sulphur--modified step barrier in  
Ref.~\cite{Xia95} is  reasonably consistent with the starting 
assumption in this case.


In this Letter, we have presented a theoretical analysis of collective 
diffusion on stepped substrates, and studied the dependence of 
the diffusion constant on the terrace width, coverage and temperature.  
In particular, we have established a criterion for crossover from 
terrace dominated diffusion to step dominated diffusion.  Compared 
with a heuristic criterion used before \cite{Xia92,Xia95}, 
our theory predicts a much stronger influence of the steps on the 
diffusion barrier.  The experiments of CO on Ni(110) 
in Refs.~\cite{Xia92,Xia93,Xia95} represent a situation characteristic of 
diffusion on smooth metal surfaces with steps as far as few hundred 
lattice spacings apart and temperature range of 100 to 200 K.
Our results show that under these conditions, 
the influence of the steps on the 
measured diffusion barrier can be considerable, and a careful theoretical 
analysis is needed to interprete the measured barrier value.
In this study, the only interaction effect taken into 
account is the exclusion of double occupation on the same site.  The 
effect of additional interaction between adatoms on the influence of 
the steps remains to be investigated.


This work has been supported by the Academy of Finland (J. M.),
Emil Aaltonen Foundation (J. M.),
and  by a grant from the Office of Naval Research (S. C. Y. and J. M.).
Computational resources of the Theoretical Physics Computing Facility
at Brown University are gratefully acknowledged.



\eject

\eject

\begin{references}

\bibitem[*]{PermAddr}
Permanent address:
Department of Physics, University of Jyv\"askyl\"a, 
P.O.\ Box 35, FIN--40351 Jyv\"askyl\"a, Finland.

\bibitem{GrowthOrdering}
M. G. Lagally (editor) {\it Kinetics of Ordering and Growth at 
Surfaces},
(Plenum, New York, 1990).

\bibitem{Sch69}
R. L. Schwoebel,
J. Appl. Phys. {\bf 44}, 614 (1969).

\bibitem{Gom90}
R. Gomer, Rep. Prog. Phys. {\bf 53}, 917 (1990).

\bibitem{Xia92}
X. D. Xiao, X. D. Zhu, W. Daum,
and Y. R. Shen, Phys. Rev. B {\bf 46}, 9732 (1992).

\bibitem{Xia93}
X. D. Xiao, Y. Xie, and Y. R. Shen,
Phys. Rev. B {\bf 48}, 17452 (1993).

\bibitem{Xia95}
X. D. Xiao, Y. Xie, C. Jakobsen, H. Galloway, M. Salmeron, and Y. R. 
Shen,
Phys. Rev. Lett. {\bf 74}, 3860 (1995).

\bibitem{Nat93}
A. Natori and R. W. Godby,
Phys. Rev. B {\bf 47}, 15816 (1993).

\bibitem{Ueb94}
C. Uebing and R. Gomer,
Surf. Sci. {\bf 306}, 419 (1994); 
{\it ibid.} 427;
{\bf 317}, 165 (1994).

\bibitem{Tah83}
R. A. Tahir--Kheli and R. J. Eliott,
Phys. Rev. B {\bf 27}, 844 (1983).

\bibitem{Ala91}
T. Ala--Nissila, J. Kjoll, S. C. Ying, and R. A. Tahir--Kheli,
Phys. Rev. B {\bf 44}, 2122 (1991).

\bibitem{CollectiveGreen}
Our rate equations and the Green's functions as defined here
correspond to collective diffusion, which can be defined solely in 
terms of density fluctuations, and thus the identity of individual
particles can be neglected \protect\cite{Ala91}.

\bibitem{Mer96}
J. Merikoski and S. C. Ying, to be published.

\bibitem{Keh84}
K. Kehr,
in {\it Applications of the Monte Carlo Method in Statistical 
Physics},
edited by K. Binder (Springer, Berlin, 1984).

\bibitem{ExactLimits}
For $c \to 0$ and $c \to 1$, the result of 
Eq.~(\protect\ref{eq:DxxDyy}) is 
exact, and in the single--particle limit $c\to 0$ it reduces to
that given in Ref.~\cite{Nat93}.

\bibitem{OtherLimits}
Other limits of $D_{xx}$ as well as those of $D_{yy}$
will be discussed in detail in forthcoming work \protect\cite{Mer96}.

\bibitem{FeiPC}
For this system the Schwoebel barrier is believed to be very small;
P. J. Feibelman, private communication.

\bibitem{Prefactors}
In case of different prefactors for jumps on terraces and at step 
edges,
$\nu_t$ and $\nu_e$, respectively, 
the effect of prefactors on crossover can be taken into account by 
using
the form $L \approx c e^{(E_B+E_S)/kT}\nu_t/\nu_e$,
from which we see that the effect of the prefactors on the crossover
temperature is only logarithmic.

\end{references}
\end{document}